\documentclass[twocolumn,superscriptaddress,amsmath,amssymb,aps,prl,floatfix]{revtex4-2}

\usepackage[usenames,dvipsnames]{color}
\usepackage{xcolor}
\usepackage{multibib} 

\usepackage{comment}
\usepackage{cancel}
\usepackage{ulem}
\usepackage{physics}
\usepackage{graphicx}
\usepackage{dcolumn}
\usepackage{multirow}
\usepackage{bm}
\usepackage{gensymb}
\usepackage{tensor}
\usepackage[percent]{overpic}
\usepackage{tikz}
\usepackage{pgfplots}
\pgfplotsset{compat=1.15}
\usepackage{mathrsfs}
\usetikzlibrary{arrows}
\definecolor{qqttcc}{rgb}{0,0.2,0.8}
\definecolor{ccqqqq}{rgb}{0.8,0,0}
\definecolor{uuuuuu}{rgb}{0.26666666666666666,0.26666666666666666,0.26666666666666666}

\usepackage[breaklinks=true,colorlinks,citecolor=blue,linkcolor=blue,urlcolor=blue]{hyperref}
\definecolor{1}{HTML}{386CB0}
\definecolor{2}{HTML}{2CA02C} 

\begin{document}
%
\title{Theory of nonlinear spin transport in  chiral conductors}
\date{\today}
\author{Lorenzo Cavicchi}
\affiliation{Scuola Normale Superiore, I-56126 Pisa,~Italy}
\author{Marco Polini}
\affiliation{Dipartimento di Fisica dell'Universit\`a di Pisa, Largo Bruno Pontecorvo 3, I-56127 Pisa, Italy}
\begin{abstract}
The chirality-induced spin selectivity (CISS) effect, discovered by Naaman and collaborators in 1999, describes the emergence of a finite spin polarization in response to current flow through a chiral electronic system. While extensive experimental studies have verified the presence of CISS in molecular systems and, more recently, in chiral materials, a complete microscopic understanding of this effect remains elusive. In this work, we propose a theoretical framework linking the  CISS effect to the orbital Edelstein effect. In the latter, a drive current induces a finite orbital magnetization, even in the absence of spin-orbit coupling. Our non-equilibrium theory naturally explains key features of the CISS effect:  its persistence in systems with weak or vanishingly small spin-orbit coupling and its connection to natural optical activity, a distinctive signature of chiral systems.
\end{abstract}

\maketitle

{\it {\color{blue} Introduction.}}---The chirality-induced spin selectivity (CISS) effect is a phenomenon wherein non-magnetic chiral structures---i.e., systems that are non-superimposable on their mirror images---allow for a preferential transmission of electrons with a specific spin orientation~\cite{naaman_review_2024,naaman_review_2020}. This effect, first observed in molecular systems~\cite{ray_science_1999, wei_jphyschem_2006} and more recently in chiral materials~\cite{Waldeck_APL_2021,rana_JACS_2025} and bulk chiral crystals~\cite{Shiota2021,Calavalle2022}, challenges conventional theories of spin-dependent electron transport, which traditionally require strong magnetic interactions. 

Numerous experiments have validated the CISS effect. Photoemission~\cite{gohler_science_2011,kettner_jphyschemC_2015,kettner_jphyschemlett_2018,privitera_chemsci_2022,bangruwa_jchemphys_2023} and electron transport~\cite{xie_nanolett_2011, kiran_advmat_2016,kiran_jchemphys_2017, mishra_jphyschemC_2020, liu_acsnano_2020, Giaconi2023} studies have confirmed that chiral molecules, such as proteins and DNA, induce exceptionally high spin polarization in transmitted electrons~\cite{al-bustami_nanolett_2020}. Scanning tunneling microscopy~\cite{aragones_small_2017} has also revealed spin-dependent conductance in single-molecule junctions. The CISS effect has broad implications across multiple disciplines. In spintronics, it provides a pathway for spin manipulation without external magnetic fields, enabling low-power spintronic devices~\cite{michaeli_jpcondmatt_2017,yang_natrevphys_2021,al-bustami_small_2018}. It also plays a role in molecular enantio-discrimination~\cite{ghosh_science_2018,lum_jphyschemlett_2021}, chemical processes~\cite{rosenberg_PRL_2008, rosenberg_angewchem_2015} and reactions~\cite{bloom_pccp_2020}, and even biology, where it may influence electron transfer processes~\cite{wei_jphyschemB_2023} and contribute to the origin of homochirality in life~\cite{michaeli_chesocrev_2016,ozturk_pnas_2022}.

Despite significant progress, the microscopic origins of the CISS effect remain debated, with various proposed mechanisms at its foundation~\cite{naaman_review_2024}. The theoretical community is primarily divided into two schools of thought. The first one considers models for chiral molecules that incorporate both chiral symmetry and spin-orbit coupling (SOC)~\cite{guo_PRL_2012, guo_2012, diniz_prl_2012, rai_jphyschem_2013,behnia_jphyschemC_2016}. While these models predict spin-dependent transport, the resulting spin polarization is usually small due to the small value of SOC in biomolecules and proteins~\cite{Evers2022}. In order to overcome the limitation of small atomic SOC in organic molecules, a re-examination of the origins of SOC in these systems led to the concept of ``geometric SOC''~\cite{gutierrez_PRB_2012, guo_PRB_2012, gutierrez_jphyschemC_2013,ghazaryan_jphyschemC_2020}, which gives rise to significantly higher magnitudes of spin polarization. 
The second class of approaches attributes spin selectivity to interactions at the chiral molecule/substrate interface~\cite{alwan_jamchemsoc_2021, Liu_NatureMater_2021, dubi_chemsci_2022,gobel_PRR_2025, sarkar_ACSnano_2025} and assumes that the chiral system is in contact with metallic leads, typically with large SOC. This mechanism certainly contributes to spin filtering in conduction measurements~\cite{adhikari_natcomm_2023}. However, photoemission experiments on helicene monolayers on different metallic substrates do not show large differences in spin polarizations~\cite{kettner_jphyschemlett_2018}.

The role of other factors, such as geometric phases~\cite{Liu_NatureMater_2021, teh_PRB_2022}, electron correlations~\cite{Fransson_JPCL_2019, chiesa_nanolett_2024, fransson_JPCL_2025}, electron-phonon coupling~\cite{Fransson2020, sang_JPCC_2021, das_jphyschemC_2022}, dissipation~\cite{Volosniev_PRB_2021}, and interaction with magnetic substrates~\cite{fransson_2024}, has been studied but is not yet fully understood. While theoretical efforts remain partially inconclusive, experimental studies have identified key characteristics common to all CISS observations~\cite{naaman_review_2024,naaman_review_2020}: (i) In transport measurements~\cite{xie_nanolett_2011}, the polarization of spin currents through chiral molecules exhibits a nonlinear dependence on the applied voltage; (ii) The spin polarization for current flowing through chiral molecules can reach exceptionally high values ($>85\%$) at room temperature, even in systems with weak or vanishingly
small SOC; (iii) The magnitude and sign of a molecule's CISS strength are directly proportional to the magnitude of its chiro-optical response (i.e.~natural optical activity)~\cite{clever_isrjchem_2022}.

One thing is certain from the theoretical point of view: in the linear response regime, Onsager-B\"uttiker reciprocity~\cite{Buttiker1988} forbids extracting spin currents from any two-terminal electrical measurement, even in the presence of auxiliary ferromagnets~\cite{Yang2019,footnote_ciss_Smatrix}. Equivalently, in a two-terminal device operated in the linear regime, the spin polarization of the flowing current cannot be accessed. A faithful description of the CISS effect in such geometries, therefore, requires going beyond the linear response regime. 

In this work, in order to transcend the linear response regime, we imagine that a chiral system is subject to a sufficiently strong electric field $\bm E$. We assume that due to electron-electron interactions, the system reaches, on a short timescale~\cite{Gantmakher1987,Bistritzer2010,sabbaghi_prb_2015,VanDuppen2016,sabbaghi_prb_2018}, a quasi-equilibrium state carrying a finite, time-independent, and spatially homogeneous current $\bm j_{\rm d}(\bm E)$. (This state will be dubbed below a {\it current-carrying state}.) The key point is that, in response to this drift current, a chiral system develops a finite spin polarization, even in the total absence of a microscopic SOC term in its Hamiltonian. The reason is that chiral systems display the so-called ``orbital Edelstein effect'' (OEE), whereby an orbital magnetization $\bm M = \bm{M}(\bm j_{\rm d}(\bm E))$ is induced by current flow~\cite{johansson_jphyscondmatt_2024, koretsune_PRB_2012, yoda_scirep_2015, yoda_nanolett_2018}. In turn, this orbital magnetization couples in a Zeeman-like fashion (i.e.~as an effective magnetic field) to the electron's spin, thereby producing a spin polarization. As we will see below, this spin polarization manifests itself through the appearance of spin-polarized currents that are non-linear functions of $\bm E$. Finally, since the OEE is related to natural optical activity~\cite{wang_PRB_2020,mahon_PRR_2020,duff_PRB_2022}, our approach allows us to also establish a direct link between the CISS effect and chiro-optical properties of chiral systems.

{\it {\color{blue} Current-carrying state of a chiral system.}}---Momentum-space occupation numbers of a driven electron system can, in principle, be obtained by solving the Boltzmann transport equation with suitable collision integrals \cite{Ziman1960,LifshitzPitaevskiiKinetics,AshcroftMermin}. Here, however, we employ a simplified phenomenological description that is often used in the literature, i.e.~the boosted Fermi surface model~\cite{GantmakherLevinson,BistritzerMacDonald2009,LucasFong2018,PoliniGeim2020,sabbaghi_prb_2015,VanDuppen2016,sabbaghi_prb_2018}. Apart from its elegance and simplicity, this model has the advantage of being exact in at least one regime, i.e.~in the hydrodynamic regime~\cite{PoliniGeim2020}, where electron-electron interactions are strong enough to produce local thermal equilibrium on a time scale $\tau_{\rm ee}\ll\tau_{\rm mr}$, where $\tau_{\rm mr}$ is the time-scale over which momentum is randomized~\cite{LucasFong2018,bauer_2015,pogna_2022}. Following Refs.~\cite{GantmakherLevinson,BistritzerMacDonald2009,sabbaghi_prb_2015,VanDuppen2016,sabbaghi_prb_2018}, we therefore assume that electron-electron collisions establish a local quasi-equilibrium state with a drift velocity $\bm v_{\rm d}$, yielding a drift current density
\begin{equation}
\bm j_{\rm d}=-en\bm v_{\rm d}~,
\end{equation}
where $-e$ is the electron charge and $n$ is the carrier density.

In turn, in a chiral system, such drift current $\bm j_{\rm d}$ generates a net orbital magnetization $\bm M$. This is the so-called OEE~\cite{johansson_jphyscondmatt_2024}---the orbital analog of the spin Edelstein effect~\cite{edelstein_1990}. In the linear response regime, the relation between the drift current and the magnetization is~\cite{yoda_scirep_2015,yoda_nanolett_2018}
\begin{equation}\label{eqn:OEE_full}
    M_\alpha(\bm q, \omega) = \int \frac{d^{D}\bm q^\prime}{(2\pi)^D} \chi_{\alpha\beta}^{\rm OE}(\bm q, \bm q^\prime, \omega) j_{\rm d,\beta}(\bm q^\prime, \omega)~,
\end{equation}
where $D$ is the dimensionality and $\chi_{\alpha\beta}^{\rm OE}(\bm q,\bm q',\omega)$ is a rank-two pseudotensor describing the OEE. In order to make analytical progress, we assume that the system is homogeneous, which leads to $\chi_{\alpha\beta}^{\rm OE}(\bm q, \bm q^\prime, \omega) = \chi_{\alpha\beta}^{\rm OE}(\bm q, \omega) \delta_{{\bm q}^\prime, -{\bm q}} $.  We then assume that, at equilibrium, the drift current is homogeneous and independent of time,~i.e.~${\bm j}_{\rm d}(\bm q, \omega)\equiv {\bm j}_{\rm d}$. This implies that, in Eq.~\eqref{eqn:OEE_full}, we can take the local ($\bm q\to\bm 0$) and static ($\omega \to 0$) limit, finding
\begin{equation}\label{eqn:OEE}
\begin{split}
M_{\alpha}&\equiv M_{\alpha}(\bm q = \bm 0, \omega =0)= \chi_{\alpha\beta}^{\rm OE}(\bm q =\bm 0,\omega =0)j_{\rm d,\beta}\\
&\equiv \chi_{\alpha\beta}^{\rm OE}j_{\rm d,\beta}~,
\end{split}
\end{equation}
where $\chi_{\alpha\beta}^{\rm OE}$ is the orbital Edelstein response, in the local and static limit. 

Since the drift current $\bm j_{\rm d}$ defines a preferential direction in space, the tensor $\chi_{\alpha\beta}^{\rm OE}$ can be further decomposed into a transverse and longitudinal component with respect to $\bm j_{\rm d}$:

\begin{equation}
\begin{split}
    \chi_{\alpha\beta}^{\rm OE} =& \chi_{\rm L}^{\rm OE}\frac{j_{\rm d,\alpha}j_{\rm d,\beta}}{j_{\rm d}^2} + \chi_{\rm T}^{\rm OE} \left(\delta_{\alpha\beta} - \frac{j_{\rm d,\alpha}j_{\rm d,\beta}}{j_{\rm d}^2} \right)~.
\end{split}
\end{equation}

With this decomposition, the constitutive relation~\eqref{eqn:OEE} reduces to
\begin{equation}\label{eqn:OEE_L}
M_\alpha= \chi_{\rm L}^{\rm OE}j_{\rm d,\alpha}~.
\end{equation}
We stress that a nonzero longitudinal OEE coefficient $\chi_{\rm L}^{\rm OE}$ requires chirality,~i.e. the absence of improper symmetry operations~\cite{flack_2003}. Indeed, the purely longitudinal isotropic relation in Eq.~\eqref{eqn:OEE_L}, corresponding to $M_\alpha \propto j_\alpha$, is symmetry-allowed only in chiral point groups. Among crystallographic point groups, this fully isotropic form is realized only in the chiral cubic classes $T$ and $O$~\cite{malgrange_book}. It is convenient to parameterize chirality by writing
\begin{equation}\label{eqn:oee_chirality}
\chi_{\rm L}^{\rm OE}=\zeta\,\tilde{\chi}_{\rm L}^{\rm OE},\qquad \text{with}~\zeta=\pm 1~.
\end{equation}

In order to study the linear response around the drifting state, we use a spin-resolved drifting Fermi distribution. We restrict our treatment to crystals within the Bloch description. The single-particle eigenstates and eigenvalues can be denoted as $|\bm k,\lambda\rangle$ and $\varepsilon_{\bm k,\lambda}$, where $\bm k$ is the (crystal) momentum, spanning the first Brillouin zone (FBZ), and $\lambda$ is the band index~\cite{AshcroftMermin}. In the present quasi-equilibrium approach, we describe the current-carrying state by a distribution function of the form
$f_{\bm k,\lambda,s}(\bm v_{\rm d}) = f\big(\varepsilon_{\bm k,\lambda} - \delta\varepsilon_{\bm k,\lambda,s}(\bm v_{\rm d})\big)$. Our expansion below only assumes that a linear-in-$\bm v_{\rm d}$ drift state exists and that the current-induced orbital magnetization produces an effective Zeeman-like splitting $\Delta_{\rm Z}(\bm v_{\rm d})$ with the opposite sign for the two spin projections.
We therefore write $\delta\varepsilon_{\bm k,\lambda,s}=\bm v_{\rm d}\cdot \bm\Pi_{\bm k,\lambda,s}$.
The precise microscopic form of $\bm\Pi_{\bm k,\lambda,s}$ depends on Galilean invariance and on band-structure details~\cite{LifshitzPitaevskiiKinetics,Narozhny2022}. 
We stress that for a Galilean-invariant band, the boosted distribution corresponds to a rigid shift of the Fermi surface. Henceforth, for concreteness, we adopt this minimal form, which captures the leading drift-induced redistribution in $\bm k$-space.

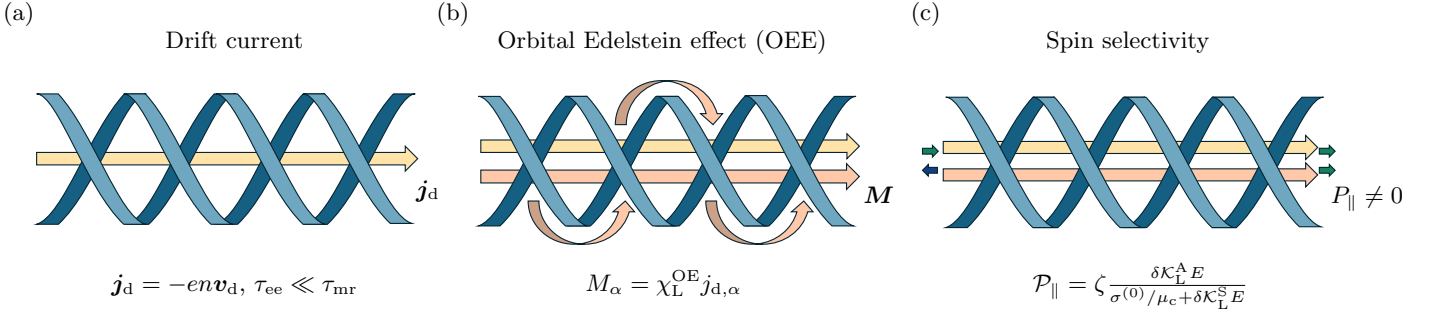
\begin{figure*}
    \centering
    \begin{overpic}[width=1\linewidth]{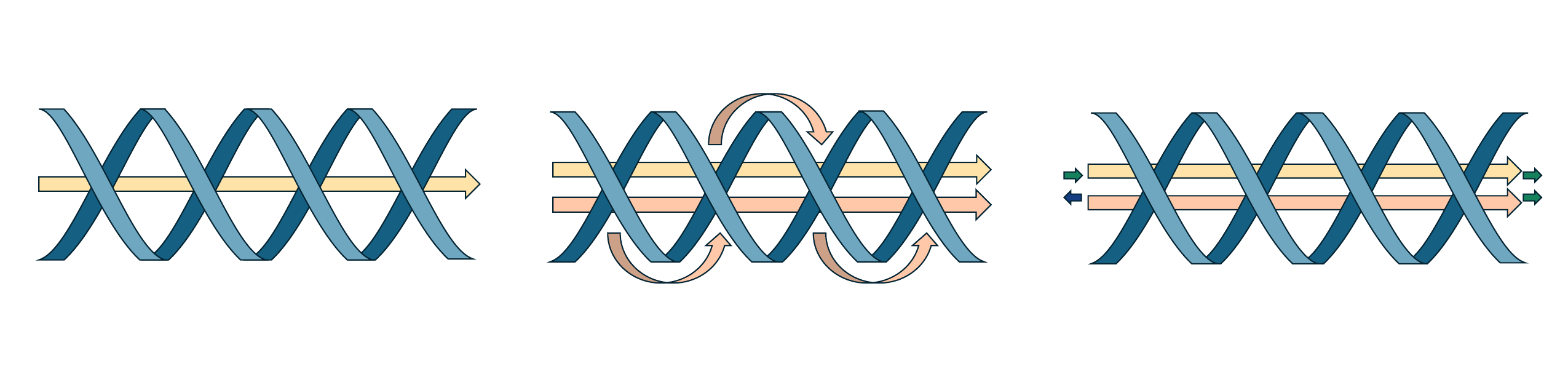}
    \put(0,22){(a)}
    \put(32,22){(b)}
    \put(67,22){(c)}
    \put(12,20){Drift current}
    \put(30.5,9){$\bm j_{\rm d}$}
    \put(8, 2){$\bm j_{\rm d} =-e n \bm v_{\rm d}$, $\tau_{\rm ee}\ll\tau_{\rm mr}$}
    \put(36.5,20){Orbital Edelstein effect (OEE)}
    \put(63.5,8.5){$\bm M$}
    \put(43, 2){$M_\alpha= \chi_{\rm L}^{\rm OE}j_{\rm d,\alpha}$}
    \put(77,20){Spin selectivity}
    \put(98,8.5){$P_\parallel\neq0$}
    \put(76,2){${\cal P}_{\parallel} = \zeta\frac{\delta{\mathcal K}_{\rm L}^{\rm A}E}{\sigma^{(0)}/\mu_{\rm c} + \delta{\mathcal K}_{\rm L}^{\rm S} E}$}
    \end{overpic}
    \caption{(a) Due to fast electron-electron interactions with respect to momentum-relaxing processes ($\tau_{\rm ee}\ll\tau_{\rm mr}$), the system reaches a quasi-equilibrium state carrying a finite, time-independent, and spatially homogeneous current $\bm j_{\rm d}$ (yellow arrow). (b) The chiral system develops an orbital magnetization $\bm{M}$ (orange arrow) due to current flow---i.e.~this is the OEE. (c) The coupling between the orbital magnetization and the electron's spin (small green and blue arrows) produces a finite spin polarization ${\cal P}_{\parallel}$.}
    \label{fig1}
\end{figure*}

The distribution of the chiral current-carrying state is therefore taken as
\begin{widetext}
\begin{equation}
\label{eqn:FD_drift}
f_{\bm k,\lambda,s=\pm}(\bm v_{\rm d})=
\frac{1}{\exp \Big\{\big[\varepsilon_{\bm k,\lambda}-\hbar\bm v_{\rm d}\cdot\bm k
- s\,\zeta\,\Delta_{\rm Z}(\bm v_{\rm d})-\mu\big]/(k_{\rm B}T)\Big\}+1}~.
\end{equation}
\end{widetext}
Here, as in Eq.~\eqref{eqn:oee_chirality}, $\zeta=\pm$ denotes the handedness of the chiral system while $s=\pm$ denotes the spin projection in the helicity basis,~i.e. with respect to the drift direction. Henceforth, for convenience, we fix the transport axis $\hat{\bm e}$, so that the drift $\bm v_{\rm d}= v_{\rm d}\cdot\hat{\bm e}$, with $v_{\rm d}$ a {\it signed} scalar.

We now write the drift-induced spin-splitting $\Delta_{\rm Z}(\bm v_{\rm d})$ as a Zeeman coupling to an effective self-induced magnetic field~\cite{footnote_magnetic_field} proportional to the longitudinal orbital magnetization, i.e.
\begin{equation}\label{eqn:DeltaZ_def}
\Delta_{\rm Z}(\bm v_{\rm d})\equiv \frac{g\mu_{\rm B}}{2}\,B_{\rm eff}(\bm v_{\rm d}),
\end{equation}
where
\begin{equation}\label{eqn:effective_B}
B_{\rm eff}(\bm v_{\rm d})=4\pi\,\lambda_{\rm M}\,\tilde M_\parallel(\bm v_{\rm d})~.
\end{equation}
In Eq.~\eqref{eqn:DeltaZ_def}, $g$ is an effective $g$-factor, and $\mu_{\rm B}$ is the Bohr magneton. The parameter $\lambda_{\rm M}$ in Eq.~\eqref{eqn:effective_B} should be understood as an effective geometry factor~\cite{JacksonEM,LandauLifshitzEDCM}.
In Eq.~(\ref{eqn:effective_B}), we introduced the chirality-even longitudinal magnetization amplitude $\tilde M_\parallel(\bm v_{\rm d})$ given by
\begin{equation}\label{eqn:m_parallel}
\tilde{M}_\parallel(\bm v_{\rm d})\equiv -en\tilde{\chi}_{\rm L}^{\rm OE}\,v_{\rm d}~.
\end{equation}
Notice that the Zeeman splitting $\Delta_{\rm Z}(\bm v_{\rm d})$ vanishes at $\bm v_{\rm d}=\bm 0$, so that the distribution \eqref{eqn:FD_drift} is spin-degenerate at equilibrium.

{\color{blue} \it Spin-resolved conductivity, spin current, and spin polarization.}---We now turn to calculate the spin-polarized currents ${\bm j}_{s}(\bm v_{\rm d})$ that emerge in response to the application of the electric field ${\bm E}$. In the DC ($\omega\tau_{\rm mr}\ll 1$) limit and to linear order in ${\bm E}$, the spin-resolved current is given by
\begin{align}\label{eqn:spin_current}
j_{\alpha,s}(\bm v_{\rm d}) &= \sigma_{\alpha \beta,s}(\bm v_{\rm d},\omega\ll 1/\tau_{\rm mr})E_\beta\nonumber\\
    &\equiv  \sigma_{\alpha \beta,s}(\bm v_{\rm d})E_\beta~.
\end{align}
Here, the local, spin-resolved AC conductivity $\sigma_{\alpha\beta,s}(\bm v_{\rm d},\omega)$ can be written quite generally as the sum of  intraband and interband contributions~\cite{giuliani_vignale_book}:
\begin{equation}\label{eq:local_cond_crystal}
\sigma_{\alpha\beta,s}(\bm v_{\rm d},\omega ) = {\cal D}_{\alpha\beta,s}(\bm v_{\rm d})\,\delta_{\tau_{\rm mr}}(\omega) + \sigma_{\alpha\beta,s}^{\rm inter}(\omega ,\bm v_{\rm d})~.
\end{equation}
Here, ${\cal D}_{\alpha\beta,s}$ is the spin-resolved Drude weight and $\delta_{\tau_{\rm mr}}(\omega)$ is a disorder-broadened delta peak,
\begin{equation}\label{eq:drude_broadening}
\delta_{\tau_{\rm mr}}(\omega)\equiv \frac{1}{\pi}\,\frac{\tau_{\rm mr}}{1+\omega^2\tau_{\rm mr}^2}~.
\end{equation}
Equivalently, the intraband response is of the usual Drude type:
$\sigma^{\rm intra}_{\alpha\beta}(\omega)=({\cal D}_{\alpha\beta,s}/\pi)\, i/(\omega+i/\tau_{\rm mr})$,
so that in the DC limit  $\sigma^{\rm intra}_{\alpha\beta}(0)=({\cal D}_{\alpha\beta,s}\tau_{\rm mr})/\pi$. The spin-resolved Drude weight is given by
\begin{align}\label{eqn:drude_weight}
&{\cal D}_{\alpha\beta,s}(\bm v_{\rm d}) =\nonumber\\
&\pi e^2 \sum_\lambda \int_{\rm FBZ} \frac{d^D\bm{k}}{(2\pi)^D}
    \Big[-f_{\bm k, \lambda,s}^\prime(\bm v_{\rm d})\Big]
    v_{\bm k\lambda,\alpha}\,v_{\bm k\lambda,\beta}~,
    \end{align}
where $v_{\bm k\lambda,\alpha}\equiv \langle \bm k,\lambda|\hat v_\alpha|\bm k,\lambda\rangle$ and $f_{\bm k, \lambda,s}^\prime(\bm v_{\rm d})$ is the first derivative with respect to energy of the chiral current-carrying distribution function $f_{\bm k, \lambda,s}(\bm v_{\rm d})$ given in Eq.~\eqref{eqn:FD_drift}. The spin-resolved interband part is
\begin{widetext}
\begin{align}\label{eqn:conductivity_inter-band}
\sigma_{\alpha\beta,s}^{\rm inter}(\omega,\bm v_{\rm d}) = \,ie^2\hbar\sum_{\lambda\neq\lambda^\prime} \int_{\rm FBZ} \frac{d^D\bm{k}}{(2\pi)^D}\left[-\frac{f_{\bm k, \lambda,s}(\bm v_{\rm d}) - f_{\bm k, \lambda^\prime,s}(\bm v_{\rm d})}{\varepsilon_{\bm k,\lambda}- \varepsilon_{\bm k,\lambda^\prime}}\right] \frac{\langle \bm k, \lambda|\hat{v}_\alpha|\bm k, \lambda^\prime\rangle \langle \bm k, \lambda^\prime| \hat{v}_\beta|\bm k, \lambda\rangle}{\hbar\omega+\varepsilon_{\bm k,\lambda} - \varepsilon_{\bm k,\lambda^\prime} + i\eta}~.
\end{align}
\end{widetext}
Here, $\eta>0$ is a phenomenological broadening (in a simple relaxation-time picture, one may take $\eta=\hbar/\tau_{\rm mr}$).

The longitudinal spin polarization ${\cal P}_{\parallel}$, used to quantify the CISS effect, is often defined as \cite{naaman_review_2024}
\begin{equation}\label{eqn:spin_polarization}
    {\cal P}_{\parallel}(\bm v_{\rm d})
    = \frac{j_{\parallel,+}(\bm v_{\rm d}) - j_{\parallel,-}(\bm v_{\rm d})}
           {j_{\parallel,+}(\bm v_{\rm d}) + j_{\parallel,-}(\bm v_{\rm d})}\,,
\end{equation}
where $j_{\parallel, s}\equiv \bm j_{s}(\bm v_{\rm d})\cdot\hat{\bm e}$ denotes the current along the transport axis, and $s=\pm$ refers to the parallel/antiparallel spin projections along the drift direction.

{\color{blue} \it Small-drift expansion.}---Since the boosted-Fermi-surface approach is controlled for small displacements, we expand $f_{\bm k,\lambda,s}(\bm v_{\rm d})$ for small $\bm v_{\rm d}$. For small drifts,
\begin{equation}\label{eqn:DeltaZ_linear}
\Delta_{\rm Z}(\bm v_{\rm d})=\mathcal Z v_{\rm d},
\qquad
\mathcal Z \equiv \left.\frac{\partial \Delta_{\rm Z}}{\partial v_{{\rm d}}}\right|_{v_{\rm d}= 0}~.
\end{equation}
Using Eq.~\eqref{eqn:DeltaZ_def}, one may write 
\begin{equation}\label{eqn:orbital_vector}
    \mathcal Z=2\pi \gamma \zeta \tilde{\chi}_{\rm L}^{\rm OE}~,
\end{equation}
with $\gamma \equiv -e n g\mu_{\rm B}\lambda_{\rm M}$.
Expanding Eq.~\eqref{eqn:FD_drift} to first order in $\bm v_{\rm d}$ yields
\begin{equation}
\begin{split}
    f_{\bm k,\lambda, s=\pm}(\bm v_{\rm d})
    &= f_{\bm k,\lambda}
    -v_{\rm d}\left(\hbar k_\parallel \pm \zeta\,\mathcal Z\right)\,f_{\bm k,\lambda}^\prime
    +{\cal O}(v_{\rm d}^2)~,
\end{split}
\end{equation}
and
\begin{widetext}
\begin{equation}
\begin{split}
    f_{\bm k,\lambda, s=\pm}^\prime(\bm v_{\rm d})
    &=
    f_{\bm k,\lambda}^\prime
    + v_{\rm d}\left(\hbar k_\parallel \pm \zeta\,\mathcal Z\right)
    \frac{1}{k_{\rm B}T}\tanh{\left(\frac{\varepsilon_{\bm k,\lambda} -\mu}{2k_{\rm B }T}\right)}\,f_{\bm k,\lambda}^{\prime}
    +{\cal O}(v_{\rm d}^2)~,
\end{split}
\end{equation}
\end{widetext}
where $k_\parallel \equiv \bm k\cdot \hat{\bm e}$, $f_{\bm k,\lambda} \equiv f_{\bm k,\lambda,s}(\bm 0)$ is the equilibrium Fermi--Dirac distribution, and
\begin{equation}
    f_{\bm k,\lambda}^\prime \equiv \frac{\partial f_{\bm k,\lambda}}{\partial \varepsilon_{\bm k,\lambda}}
    = -\frac{1}{4k_{\rm B}T\cosh^2\left(\frac{\varepsilon_{\bm k,\lambda} -\mu}{2k_{\rm B }T}\right)}~.
\end{equation}

Substituting the expansions above into Eqs.~\eqref{eq:local_cond_crystal}--\eqref{eqn:conductivity_inter-band} and collecting terms up to first order in $\bm v_{\rm d}$ yields
\begin{equation}\label{eq:local_cond_expansion}
\sigma_{\alpha \beta, s=\pm}(\bm v_{\rm d})
\equiv
\sigma_{\alpha \beta}^{(0)} + \delta \sigma_{\alpha \beta}^{\rm S}(\bm v_{\rm d})
\pm \zeta\, \delta \sigma_{\alpha \beta}^{\rm A}(\bm v_{\rm d})~,
\end{equation}
where $\sigma_{\alpha\beta}^{(0)}\equiv \sigma_{\alpha\beta,s}(\bm v_{\rm d}=\bm 0)$ is the equilibrium (spin-degenerate) conductivity, evaluated in the DC limit. The first-order spin-even (S) and spin-odd (A) corrections can be written as
\begin{widetext}
\begin{equation}
\begin{split}\label{eqn:delta_sigma_S}
\delta \sigma_{\alpha \beta}^{\rm S}(\bm v_{\rm d})
&=
-\delta_{\tau_{\rm mr}}(\omega)\frac{\pi e^2}{k_{\rm B}T} \sum_\lambda \int_{\rm FBZ} \frac{d^D\bm{k}}{(2\pi)^D}
f_{\bm k, \lambda}' \tanh{\left(\frac{\varepsilon_{\bm k,\lambda} -\mu}{2k_{\rm B }T}\right)}
\,v_{\bm k\lambda,\alpha} v_{\bm k\lambda,\beta}\,(\hbar k_\parallel v_{\rm d})
\\
&\quad
+\,ie^2 \hbar \sum_{\lambda\neq \lambda^\prime}\int_{\rm FBZ}\frac{d^D\bm k}{(2\pi)^D}
\left[\frac{f_{\bm k,\lambda}'-f_{\bm k,\lambda^\prime}'}{\varepsilon_{\bm k, \lambda}-\varepsilon_{\bm k, \lambda^\prime}}\right]
\frac{\langle \bm k, \lambda|\hat{v}_{\alpha}|\bm k, \lambda^\prime\rangle\langle \bm k, \lambda^\prime|\hat{v}_{\beta}|\bm k, \lambda \rangle}{\varepsilon_{\bm k, \lambda}-\varepsilon_{\bm k, \lambda^\prime} + i\eta}\,
(\hbar k_\parallel v_{\rm d})
\\
&\equiv \delta{\mathcal K}_{\alpha\beta}^{\rm S} v_{\rm d}~,
\end{split}
\end{equation}
\begin{equation}
\begin{split}\label{eqn:delta_sigma_A}
\delta \sigma_{\alpha \beta}^{\rm A}(\bm v_{\rm d})
&=
-\delta_{\tau_{\rm mr}}(\omega)\frac{\pi e^2}{k_{\rm B}T} \sum_\lambda \int_{\rm FBZ} \frac{d^D\bm{k}}{(2\pi)^D}
f_{\bm k, \lambda}' \tanh{\left(\frac{\varepsilon_{\bm k,\lambda} -\mu}{2k_{\rm B }T}\right)}
\,v_{\bm k\lambda,\alpha} v_{\bm k\lambda,\beta}\,(\mathcal Z v_{\rm d})
\\
&\quad
+\,ie^2 \hbar \sum_{\lambda\neq \lambda^\prime}\int_{\rm FBZ}\frac{d^D\bm k}{(2\pi)^D}
\left[\frac{f_{\bm k,\lambda}'-f_{\bm k,\lambda^\prime}'}{\varepsilon_{\bm k, \lambda}-\varepsilon_{\bm k, \lambda^\prime}}\right]
\frac{\langle \bm k, \lambda|\hat{v}_{\alpha}|\bm k, \lambda^\prime\rangle\langle \bm k , \lambda^\prime|\hat{v}_{\beta}|\bm k, \lambda \rangle}{\varepsilon_{\bm k, \lambda}-\varepsilon_{\bm k, \lambda^\prime}+ i\eta}\,
(\mathcal Z v_{\rm d})
\\
&\equiv \delta{\mathcal K}_{\alpha\beta}^{\rm A} v_{\rm d}~.
\end{split}
\end{equation}
\end{widetext}
In the last equalities, we used the fact that the corrections are linear in $\bm v_{\rm d}$ and can therefore be encoded by second-rank tensors $\delta{\mathcal K}^{\rm S/A}_{\alpha\beta}$.

The current density for each spin species follows from
\begin{equation}\label{eqn:spin_current2}
\begin{split}
    j_{\alpha,s=\pm} &= \sigma_{\alpha \beta,s =\pm}(\bm v_{\rm d},\omega=0)\,E_{\beta}\\
    &= \sigma^{(0)}_{\alpha\beta}E_{\beta}
    + \delta{\mathcal K}_{\alpha\beta}^{\rm S} E_{\beta}v_{\rm d} 
    \pm \zeta\,\delta{\mathcal K}_{\alpha\beta}^{\rm A} E_{\beta}v_{\rm d}.
\end{split}
\end{equation}

Under bias reversal $\bm E\to -\bm E$, the steady drift reverses as $\bm v_{\rm d}\to -\bm v_{\rm d}$. 
The spin-selective part of the current follows directly from Eq.~\eqref{eqn:spin_current2}:
$\Delta j_\alpha \equiv j_{\alpha,+}-j_{\alpha,-}
=2\zeta\,\delta\mathcal K^{\rm A}_{\alpha\beta}\,E_\beta v_{\rm d}$, 
which is invariant under $(\bm E,\bm v_{\rm d})\to(-\bm E,-\bm v_{\rm d})$ because it depends on the even product $v_{{\rm d}}E_\beta$. Consequently, the helicity-resolved current difference is even in bias. By contrast, when the current is projected onto a fixed laboratory spin basis, the relative ordering of the two spin-resolved current channels changes under bias reversal because the mapping between laboratory spin and the helicity labels $s=\pm$ flips when the drift direction reverses~\cite{footnote_labframe}. 

We now specialize to a chiral and isotropic conductor driven along a fixed transport axis, so that $\sigma^{(0)}_{\alpha\beta}=\sigma^{(0)}\delta_{\alpha\beta}$ and the drift is longitudinal. Phenomenologically, the steady drift velocity is related to the applied field by the carrier mobility $\mu_{\rm c}$,
\begin{equation}
    v_{\rm d} =\mu_{\rm c}E.
\end{equation}
Projecting Eq.~\eqref{eqn:spin_current2} onto the transport axis yields~\cite{footnote_mobility}
\begin{equation}\label{eqn:spin_current_non_linear}
\begin{split}
    j_{\parallel,s=\pm} = \sigma^{(0)} E + \mu_{\rm c}\left(\delta{\mathcal K}_{\rm L}^{\rm S}  \pm \zeta\, \delta{\mathcal K}_{\rm L}^{\rm A}\right)E^2~,
\end{split}
\end{equation}
where $\delta{\mathcal K}_{\rm L}^{\rm S/A}\equiv \delta{\mathcal K}_{\alpha\beta}^{\rm S/A}\hat e_\alpha\hat e_\beta$ is the longitudinal projection along the transport axis $\hat{\bm e}$. From Eq.~\eqref{eqn:spin_current_non_linear}, a nonlinear spin-selective contribution arises once a finite drift is established, consistent with what is commonly reported in CISS transport measurements~\cite{xie_nanolett_2011,kiran_advmat_2016,kiran_jchemphys_2017,mishra_jphyschemC_2020,liu_acsnano_2020,Giaconi2023}. 

Finally, the spin polarization \eqref{eqn:spin_polarization} becomes
\begin{equation}\label{eqn:P_parallel_final}
   {\cal P}_{\parallel} = \zeta\frac{\delta{\mathcal K}_{\rm L}^{\rm A}E}{\sigma^{(0)}/\mu_{\rm c} + \delta{\mathcal K}_{\rm L}^{\rm S} E}~.
\end{equation}
Equation \eqref{eqn:P_parallel_final} makes explicit that ${\cal P}_\parallel$ is controlled by the drift-induced spin splitting through $\delta{\mathcal K}_{\rm L}^{\rm A}$, which depends on $\mathcal Z$, and, via Eq~\eqref{eqn:orbital_vector}, on the longitudinal orbital Edelstein response $\chi_{\rm L}^{\rm OE}=\zeta\tilde{\chi}_{\rm L}^{\rm OE}$. In this framework, a larger OEE implies a larger CISS spin polarization, with the chirality dependence encoded by $\zeta=\pm$.

{\it {\color{blue} Connection to natural optical activity.}}---We conclude by establishing a connection between the CISS effect and a distinct property of chiral systems, i.e.~natural optical activity. 

We first observe that the OEE and natural optical activity in chiral systems are profoundly related~\cite{wang_PRB_2020,mahon_PRR_2020,duff_PRB_2022,MaPesin2015,Zhong2016,Sahin2018,Hara2020}. This relation can be understood by following energy considerations~\cite{pershan_PR_1963}. All magneto-electric effects can be derived from the thermodynamics of a chiral system subjected to an electromagnetic (EM) field $(\bm E(\omega), \bm H(\omega))$. Neglecting higher magnetic multipoles other than the magnetic dipole, the simplest free-energy contribution that describes magneto-electric effects is~\cite{pershan_PR_1963}:
\begin{equation}\label{eqn:free_energy_magnetoelectric}
\begin{split}
    {\cal F}(\bm E, \bm H) &= -\left[\chi_{\alpha\beta}(\omega)E_\alpha^*(\omega)H_\beta(\omega) + \chi_{\alpha\beta}^*E_\alpha(\omega)H_\beta^*(\omega)\right].
\end{split}
\end{equation}
This expression for the free energy must be complemented with the constitutive relations for polarization $\bm P(\omega)$ and magnetization $\bm M(\omega)$:

\begin{gather}
    \label{eq:polarization}
    P_\alpha(\omega) = -\frac{\partial {\cal F}}{\partial E_\alpha^*(\omega)}\bigg|_{\bm E = \bm 0}~,\\
    \label{eq:magnetization}
    M_\alpha(\omega) = -\frac{\partial {\cal F}}{\partial H_\alpha^*(\omega)}\bigg|_{\bm H = \bm 0}~.
\end{gather}
In a time-reversal invariant crystal, we have $\chi_{\alpha\beta}(\omega) = -\chi_{\alpha\beta}^*(\omega)$. From Maxwell equations, one can quantify the contribution to the dielectric tensor $\epsilon_{\alpha\beta}$ induced by the free energy of Eq.~\eqref{eqn:free_energy_magnetoelectric}. Using the fact that the crystal symmetries are high enough that $\chi_{\alpha\beta}(\omega) = \chi_{\alpha\alpha}(\omega)\delta_{\alpha\beta}$, the contribution to the dielectric tensor is~\cite{pershan_PR_1963}
\begin{equation}\label{eqn:optical_activity}
    \epsilon_{\alpha\beta}(\omega) = \delta_{\alpha\beta} + 4\pi  \varepsilon_{\alpha\gamma\beta}\frac{q_\gamma c}{\omega} \left(\chi_{\alpha\alpha}(\omega) + \chi_{\beta\beta}^*(\omega)\right)~.
\end{equation}
Eq.~\eqref{eqn:optical_activity} describes what is known as \textit{natural optical activity} (NOA)~\cite{pershan_PR_1963,LandauVIII}.  We emphasize that, from the constitutive relation~\eqref{eq:magnetization}, the electric field generates a magnetization:
\begin{equation}\label{eqn:GME}
    M_\alpha(\omega ) = \chi_{\beta\alpha}^*(\omega)E_\beta(\omega)~.
\end{equation}
This is called the \textit{magnetoelectric effect} (MEE)~\cite{MaPesin2015,vanderbilt,fiebig_2015,malashevich_2010}. Therefore, the NOA described by the dielectric tensor~\eqref{eqn:optical_activity} and the MEE~\eqref{eqn:GME} arise from the same magnetoelectric mechanism encoded in $\chi_{\alpha\beta}(\omega)$. Finally, we make contact with the constitutive relation for the OEE~\eqref{eqn:OEE}, relating the MEE to the OEE~\cite{yoda_nanolett_2018}:
\begin{equation}\label{eqn:OEE_OA}
    \chi_{\beta\alpha}^*(\omega) = \chi_{\alpha\gamma}^{\rm OE}(\omega)\sigma^{(0)}_{\gamma\beta}(\omega)~,
\end{equation}
where $\sigma^{(0)}_{\alpha\beta}(\omega)$ is the spin-degenerate equilibrium optical conductivity evaluated at finite frequencies.
Eq.~\eqref{eqn:OEE_OA} is a very important relation that explicitly connects the NOA with the OEE. Within our theoretical framework, this link provides a direct proportionality between the CISS strength and the chiro-optical response in molecules and chiral crystals.

{\it {\color{blue} Conclusions.}}---We proposed a macroscopic, internally consistent picture of the CISS effect based on quasi-equilibrium transport in a chiral current-carrying state. In our approach, a drift current induces, via the OEE, an orbital magnetization $\bm M$. The electron spin couples to $\bm M$, favoring transport toward one spin orientation and producing spin-polarized current. We showed how this interpretation of the CISS effect naturally describes a
non-linear CISS effect that persists in the limit of very weak SOC, as OEE manifests itself without the need for SOC. Furthermore, we related the CISS magnitude to the NOA of the chiral system. More broadly, since our theory is based on linear response (although around a chiral current-carrying state), it provides a powerful basis to assess how electron correlations, quantum geometry, and other microscopic mechanisms influence the OEE response, the spin conductivity, and, ultimately, the CISS effect. Furthermore, since our calculation is based on a chiral current-carrying state, it can also be applied to non-transport set-ups in which photo-induced electrons are injected into the chiral system. 
We believe that the interplay between the OEE and other fundamental mechanisms in chiral materials, such as geometric SOC and electron correlations, will provide a comprehensive picture of the CISS effect, potentially bridging the gap between theoretically and experimentally measured spin polarizations. 

We conclude by noting that recent experiments have reported a finite spin polarization already in the linear-response regime, an apparent violation of Onsager reciprocity that remains to be understood (see discussion in Ref.~\cite{binghai_2024} and references therein).

\section{Acknowledgments}
M. P. wishes to thank Claudia Felser for introducing him to the CISS effect and Aharon Kapitulnik, Frank Koppens, and Binghai Yan for many interesting discussions. L. C. thanks Riccardo Bertini, Francesco Cioni, and Guido Menichetti for insightful discussions and for critically reading the manuscript. L. C. and M. P. were supported by the European Union under grant agreement No. 101131579 - Exqiral. Views and opinions expressed are however those of the authors only and do not necessarily reflect those of the European Union or the European Commission. Neither the European Union nor the granting authority can be held responsible for them.


\begin{thebibliography}{99}
\bibitem{footnote_labframe}
To compare with experiments using a fixed ferromagnetic axis, let $\rightarrow,\leftarrow$ denote spin states defined in the laboratory frame. If $\rightarrow$ coincides with $s=+$ for $E>0$, then for $E<0$ the same laboratory state corresponds to $s=-$, because $\hat{\bm v}_{\rm d}$ reverses. Hence
\begin{equation}
j_{\rightarrow}(E)=
\begin{cases}
j_{+}(E), & E>0,\\
j_{-}(E), & E<0,
\end{cases}\nonumber
\end{equation}
and
\begin{equation}
j_{\leftarrow}(E)=
\begin{cases}
j_{-}(E), & E>0,\\
j_{+}(E), & E<0.
\end{cases}\nonumber
\end{equation}

For $j_{\pm}(E)=\sigma^{(0)}E+\mu_c(\delta K_L^S\pm \zeta\delta K_L^A)E^2$, one then finds
\begin{equation}
j_{\rightarrow}-j_{\leftarrow}=2\mu_c\zeta\delta K_L^A E|E|~,\nonumber
\end{equation}
so that for $\mu_c\delta K_L^A>0$ one has $j_{\rightarrow}>j_{\leftarrow}>0$ at $E>0$ and $j_{\rightarrow}<j_{\leftarrow}<0$ at $E<0$ for positive chirality ($\zeta = +$).
\bibitem{footnote_mobility}
In writing Eq.~\eqref{eqn:spin_current_non_linear}, we have kept the lowest-order phenomenological relation $v_{\rm d}=\mu_{\rm c}E$, so that the leading nonlinearity originates from the drift-induced modification of the conductivity. Additional nonlinearities in $v_{\rm d}(E)$ can be incorporated straightforwardly but are beyond the minimal theory presented here.
\end{thebibliography}
\end{document}